\documentclass[12pt,preprint]{aastex}

\begin{document}
\title{Effects of neutrino-driven kicks on the supernova
explosion mechanism}

\author{Christopher L. Fryer\altaffilmark{1,2} and Alexander 
Kusenko\altaffilmark{3} }

\altaffiltext {1}{Department of Physics, The University of Arizona,
  Tucson, AZ 85721} 
\altaffiltext{2}{Theoretical Division, LANL, Los Alamos, NM 87545}
\altaffiltext{3}{Department of Physics and Astronomy, UCLA, Los
  Angeles, CA 90095-1547}

\begin{abstract}

We show that neutrino-driven pulsar kicks can increase the energy of
the supernova shock. The observed large velocities of pulsars are
believed to originate in the supernova explosion, either from
asymmetries in the ejecta or from an anisotropic emission of neutrinos
(or other light particles) from the cooling neutron star. In this
paper we assume the velocities are caused by anisotropic neutrino
emission and study the effects of these neutrino-driven kicks on the
supernova explosion.  We find that if the collapsed star is marginally
unable to produce an explosion, the neutrino-driven mechanisms can
drive the convection to make a successful explosion.  The resultant
explosion is asymmetric, with the strongest ejecta motion roughly in
the direction of the neutron star kick.  This is in sharp contrast
with the ejecta-driven mechanisms, which predict the motion of the
ejecta in the opposite direction.  This difference can be used to
distinguish between the two mechanisms based on the observations of
the supernova remnants.

%

\end{abstract}


\section{Introduction}

Current observations of pulsar proper motions suggest that a large
fraction of neutron stars are moving with velocities in excess of
400\,km\,s$^{-1}$ \citep{Cor98,Fry98,Lai01,Arz02}.  The large energy
and momentum released during the formation of the neutron star (and
the ensuing supernova explosion), coupled with the growing evidence
that many core-collapse supernovae exhibit asymmetric explosions, has
led to a general consensus in astronomy that neutron stars receive
these large ``kicks'' at birth.  The mechanisms driving these kicks
can be separated into two classes: ejecta-driven kicks and the kicks driven by
emission of neutrinos or other weakly interacting particles.  

Ejecta-driven kicks can occur if a sufficient degree of anisotropy
develops in the hydrodynamics of the explosion.  Since only 1\% of the
collapse energy accompanies the ejecta, large asymmetries are required
to produce large supernova kicks.  A number of ejecta asymmetries have
been proposed: asymmetric collapse \citep{Bur96}, low mode convection
\citep{Her92,Bur03} and the related low-mode convection in an
accretion shock instability \citep{Blo03}.  Asymmetries in the
progenitor star can not produce kicks in excess of 200\,km\,s$^{-1}$
\citep{Fry04}, far short of the observed 1000\,km\,s$^{-1}$.
Asymmetries produced by low mode convection has proven more successful
\citep{Bur03} in 2-dimensional studies.  Such mechanisms require, by
momentum conservation, that the kick be along the explosion asymmetry,
but moving in the opposite direction of the ejecta.

Asymmetric neutrino emission has been proposed as an alternate kick
mechanism.  This mechanism takes advantage of the fact that most of
the energy and momentum released in the collapse of a massive star is
in the form of neutrinos, and asymmetries of a percent are sufficient
to produce the observed kicks.  The proposed mechanisms range from
collective effects, for example, turbulence near the neutrinosphere
\citep{Soc05}, to elementary processes involving neutrinos, including
neutrino oscillations \citep{Kus96,Kus97,Bar02,Ful03,Bar04,Kus04}.
All these mechanisms require strong magnetic fields. Although the
surface magnetic fields of ordinary radio pulsars are estimated to be
of the order of $10^{12}-10^{13}$G, the magnetic field inside a
neutron star may be much higher, probably as high as $10^{16}$G
\citep{magnetic,dt,magnetars}.

Naively, one might think that even the standard urca reactions responsible for
production of neutrinos, $ p+e^-\rightleftharpoons \nu_e+n$ and
$\bar\nu_e+p\rightleftharpoons n+e^+$, have a sufficient asymmetry to give the
neutron star a kick.  Indeed, in the rates of the urca processes depend on the 
relative orientations of the electron spins and the neutrino momentum.  Hence,
there is a 10-20\% anisotropy in the distribution of
neutrinos in every one of these processes~\citep{chugai,drt}. However, this
asymmetry in production does not lead to any asymmetry in the emission of
neutrinos, because the anisotropy is washed out by the re-scattering of
neutrinos on their way out of the star \citep{Vil95,Kus98,Arr99}.  If some
other particles, with interactions weaker than those of neutrinos were produced
anisotropically, their emission would remain anisotropic.  For example, if
sterile neutrinos exist and have a small mixing with active neutrinos, they
should be produced in the urca processes at the rate suppressed by the square
of the mixing angle \citep{Ful03}.  It is intriguing that the parameters
of the sterile neutrinos required for the pulsar kicks \citep{Kus97,Ful03}
are consistent with the mass and mixing that make the sterile neutrino a good
dark matter candidate \citep{Ful03}.

There is a strong evidence that most of the gravitating matter in the universe
is not made of ordinary atoms.  This evidence is based on a consensus of
observations of galaxy rotation curves, cosmic microwave background
radiation, gravitational lensing, and X-ray emission from galaxy clusters. 
None of the known particles can be the dark matter, and a number of candidates
have been proposed.  Perhaps, the simplest extention of the Standard Model
that makes it consistent with cosmology is the addition of a sterile neutrino
with a 2-15 keV mass.  Unlike the active fermions, which must be added in the
whole generations to satisfy the anomaly constraints, or the supersymmetric
particles, which require a major modification of the particle content, the
sterile neutrino does not entail and additional counterparts because it is
gauge singlet. Sterile neutrinos can be produced from
neutrino oscillations in the early universe in just the right amount to be the
dark matter \citep{dw,Aba01,Aba01a,Dol02,Map05}. If their mass exceeds 2~keV,
they are sufficiently cold to explain the large-scale structure.  

The discovery of neutrino oscillations points to the existence of some gauge
singlets, at least those that make the right-handed counterparts of the active
(left-handed) neutrinos.  However, the number of \textit{sterile} neutrinos is
still unknown. Unless some neutrino experiments are wrong, the present data on
neutrino oscillations cannot be explained without sterile neutrinos. Neutrino
oscillations experiments measure the differences between the squares of
neutrino masses, and the results are: one mass squared difference is of the
order of $10^{-5}$(eV$^2$), the other one is $10^{-3}$(eV$^2$), and the third
is about $1\,$(eV$^2$).  Obviously, one needs more than three masses to get the
three different mass splittings which do not add up to zero.  Since we know
that there are only three active neutrinos, the fourth neutrino must be
sterile.  However, if the light sterile neutrinos exist, there is no compelling
reason why their number should be limited to one.  Some theoretical arguments
favor at least three sterile neutrinos \citep{Asaka:2005an}.  If there are
three sterile neutrinos, they can help explain the matter-antimatter asymmetry
of the universe \citep{Asaka:2005pn}. 

Oscillations to sterile neutrinos add an intriguing additional
consequence to the search for a neutron star kick mechanism; the
opportunity to use supernovae as laboratories to study particle
physics.  Other weakly interacting particles, for example, majorons, may cause
the asymmetry as well \citep{Far05}.  Supernova asymmetries can be used to
discover or constrain a class of weakly interacting particles with masses below
100 MeV.  It is useful, therefore, to separate the details of a particular kick
mechanism from its effects on the supernova, and to perform a model-independent
analysis of how the non-ejecta kicks impact the rest of the supernova.  This
is the main goal of the present paper. 

The neutrino-driven explosion mechanism has evolved considerably since
its introduction by \cite{Col66}.  Although it is becoming increasingly 
accepted that convection above (and possibly within) the proto-neutron
star can help make neutrino heating efficient enough to drive an
explosion, the current state-of-the-art produces a range of results
\citep{Bur95,Jan96,Mez98,Fry99,Fry00,FW02, Bur03, FW04,Wal05}.  Over
the past few years, a number of papers have studied ways to make the
convection more vigorous, from asymmetries in the collapse
\citep{Bur96,Fry04} to instabilities in the accretion shock and a possible
vortical-acoustic instability \citep{Blo03,Sch04}.  The neutrino-driven kicks
have the effect of breaking the spherical symmetry of the overall explosion,
 which may help stir the material and strengthen the convection.

In this paper, we study the effects of the neutrino-oscillation 
kick mechanism on the core-collapse engine.  We test its ability 
to help drive an explosion and study the observational implications 
of an explosion driven by the asymmetric emission from neutrinos.  
\S 2 describes our computational set-up and the results of the 
simulations.  We find that, under some conditions, neutrino-driven kicks can
affect the explosion.  In \S 3, we study 
this effect and how it aids the explosion mechanism.  We conclude 
with a discussion of the observational implications from these 
effects and how these observations constrain what we know about 
neutrino oscillations.

\section{Simulations}

All our simulations begin with the standard 15\,M$_\odot$ progenitor
``s15s7b2'' produced by \cite{WW95}.  This progenitor is then mapped
into 3-dimensions using a series of spherical shells.  The entire
evolution from bounce to explosion (if an explosion occurs) is modeled
using SNSPH \citep{FRW05}.  High resolution simulations using this
code develop convection almost immediately after bounce and produce
delayed neutrino explosions 150\,ms later.  The effect of kicks
driven by neutrino oscillations is small for these quick explosions.

For this paper, we focus on the effect these kicks can have on a
marginal explosion.  The basic picture behind the convectively driven
explosions can be described with a pressure-cooker
analogy\citep{Her94, Fry99}.  Neutrinos from the proto-neutron star
(heat source) drive convection that pushes against the ram-pressure
force of the infalling star (lid of the pressure cooker).  Where the
infalling material hits the upward flows of the convective region, an
accretion shock forms.  If the convection can drive this accretion
shock outward, an explosion occurs \citep{Fry99}.  Hence, if we delay
or damp out the convection, the star is less likely to explode and we
can turn an explosion into a fizzle.  We believe the differences in
the convective modeling explain many of the differences between the
simulations over the past decade.

We have made a series of modifications to delay convection, ultimately
causing our zero kick models to fail to produce supernova explosions.
These modifications include switching the equation of state to reduce
the initial post-bounce entropy gradient to reducing the simulation
resolution and increasing the artificial viscosity to damp out the
initial perturbations.  With these modifications, our spherically
symmetric model does not produce an explosion, and at the end of the
simulation, the shock radius has already begun to move inward
(Fig. 1).  Although this simulation does not explode, it straddles the
explosion/fizzle boundary.  Our models including strong neutrino-driven kicks
develop much more convection (Fig. 1), and as we shall see later in
this section, ultimately produce asymmetric explosions.  Before we
discuss these results, let's review the modifications to the code that
delayed the convection.

First, we effectively remove the coupled equation of state developed
in \cite{Her94} by lowering the critical density for the
Lattimer-Swesty (LS) equation of state \cite{Lat91} to $10^9 {\rm g
cm^{-3}}$ (from $10^{11} {\rm g cm^{-3}}$). A now-accepted error in
the energy levels of the Lattimer-Swesty network leads to a slightly
different density/temperature boundaries for abundance states
(e.g. the alpha particle peak: see \cite{Tim05} for details).  This
can lead to very different entropy profiles after bounce.  Figure 2
shows the entropy profile with the lowered critical density (dark
dots) versus the coupled network (light dots) 40\,ms after bounce
(models Stan0, HVisc0 from Table 1).  The vertical lines correspond to
the shock positions for the the coupled network (light) and lowered
density model (dark).  In the past, this difference in entropy profile
has been attributed to either the simplified neutrino transport scheme
of SNSPH or the inability of SNSPH to model shocks.  We now believe
that these differences are caused by equation-of-state differences.
The smooth features of the entropy across the coupled equation of
state shock occurs because nuclear burning is altering the entropy
evolution.  Although the entropy profile using the lowered critical 
density for the LS equation of state is less prone to convection, 
this alone is not enough to significantly alter the explosion.

Probably the most important modification was our use of low resolution
calculations (100,000-150,000 particles).  Convective modes roughly
below the smoothing-length size-scale will not grow \citep{Fried05}.
By coarsening our resolution, many of the initial convective seeds do
not grow and the convection is significantly delayed (Fig. 1).  We
have additionally increased the artificial viscosity of SPH while
reducing the viscous heating to minimize overheating.  The importance
of these two changes can be seen in Table 1.  The ultimate result 
of all these changes is that, for our symmetric runs, no explosion 
occurs.  To drive an explosion, we must find a way to enhance the 
convection.

One way to enhance the convection is to drive larger-scale convective
modes either by producing more global asymmetries in the density or
temperature or by weakening the ram-pressure preventing the explosion.
In the next section, we will review the role of neutrino-driven kicks
can play in modifying this above list.  Conservation of momentum
requires that the proto-neutron star react to the asymmetric emission
of sterile neutrinos, giving the neutron star a kick.  To mimic the
effect of neutrino oscillations, we have modified the SNSPH code to
include an artificial acceleration term to material that rises above a
critical density.  For our simulations, we used a critical density set
to $10^{11} {\rm g \,cm^{-3}}$.  The results do not change noticeably
for critical densities lying between $10^{10}-10^{14} {\rm g \,
cm^{-3}}$ as this material is all ultimately part of the proto-neutron
star and most of the proto-neutron star's mass is at densities above
$10^{14} {\rm g \,cm^{-3}}$.  The velocities of the proto-neutron star
using the entire range of critical densities lie within 10\% of each
other.

The two plots of the x-y plane comparing our ``kicked'' and symmetric
models shows the difference in the shocks and the structure of the
convection 92\,ms after bounce.  Figure 3 highlights the magnitude of
this difference by plotting radial velocity versus radius for these 2
models at this same time.  Figure 4 shows the high viscosity kicked
model (HVisc100) at the end of the simulation.  The neutrino-driven
kick led to an explosion with strong asymmetries in the direction of
the kick.  In the next section, we review the cause of this explosion
revival and its accompanying asymmetry.

\section{Understanding the Asymmetry}

The explosion asymmetry in Figure 4 has a number of implications for
observations of this neutrino-induced kick mechanism.  Before we
discuss these implications, we must first understand the cause of the
asymmetry.  A number of effects could lead to the asymmetry we
observe:  i) material ahead of the moving neutron star is heated more
efficiently by neutrinos diffusing out of the neutron star (or this
material is heated by the moving neutron star), ii) the ram pressure of
the accretion shock is weakened ahead of the neutron star, or iii) convection 
is driven by the motion of the neutron star (growing stronger ahead of 
the neutron star's motion).  These different options are summarized 
in Figure 5.

One can imagine two ways in which the material ahead of the moving
neutron star could possibly be heated more effectively than the
material in its wake: neutrinos deposition or shocks from the moving
neutron star.  To test the neutrino deposition, Figure 6 shows the
energy deposited by neutrinos above an assumed spherical gain region
(Fryer 2004 found that the gain radius is actually larger in the
direction of the proto-neutron star's motion).  It appears that there
is more energy deposition for the moving neutron star.  Indeed, the
energy deposition for particles with $x>0$ is $5.1 \times 10^{51} {\rm
erg s^{-1}}$ compared to the $1.8 \times 10^{51} {\rm erg s^{-1}}$ for
particles with $x<0$.  For the symmetric simulation, the energies for
particles with $x<0$ and $x<0$ are equal: $3.5 \times 10^{51} {\rm erg
s^{-1}}$.  But this estimate is misleading.  If we center the
simulation about the center of the neutron star, the energy deposition
for particles $x-x_{\rm NS}>0$ is within 1\% of the deposition for
particles with $x-x_{\rm NS}<0$.  It is unlikely that asymmetric
heating is causing the explosion asymmetry.  The kinetic energy of the
neutron star on the matter also is a minor effect; the total energy
deposited by motion of the neutron star is less than 0.01\% of the
total matter energy.  

Alternatively, the ram pressure of the infalling shock could be
diminished on the edge leading the neutron star motion.  But a study
of our shocks shows that the pressure of the accretion shock on the
leading and trailing edges do not differ by more than a few percent.
What is different is that the mass in the region between the accretion
shock and the proto-neutron star is higher in front of the shock than
behind; the effect of the neutron star's bow shock and wake as it
moves through the collapsing star.  Figure 7 shows the relative mass
in a 15$^{\circ}$ cone leading the neutron star and trailing the
neutron star as a function of radial bins out from the neutron star.
Below 300\,km, the mass is enhanced in front of the neutron star.  If
the neutron star were moving supersonically, this would be the bow
shock.  Behind the neutron star, in its wake, the density is lowered.
Beyond 300\,km, the density is lower ahead of the neutron star.  This
occurs because the neutron star is moving through the collapsing star
and is hitting the lower-density layers of the star in front of it.  

This effect helps build an explosion (and an explosion asymmetry) in
two ways.  First, the total energy in the budding convective region
(the region between the proto-neutron star and the infalling shock) is
much larger ahead of the proto-neutron star than behind.  This is
simply a restatement of the fact that the mass is piling up in front
of the proto-neutron star.  Figure 8 shows the energy distribution of
matter in a shell 70-110\,km away from the proto-neutron star as a
function of angle.  From figure 8 we clearly see the effect of this
mass pile-up.  Over 70
ahead of the proto-neutron star and it is peaked directly in front of
the neutron star's motion.  Although not a direct cause of convection,
the increased energy is conducive to turnover and ultimately mixing.
Second, because the density of the infalling material is lower ahead
of the neutron star, the shock will experience a lower pressure when
it reaches this point, making it easier to push the shock forward and
drive an explosion.  We believe these effects, the stimulation of
convection, turns the ``fizzle'' into an explosion.  However, if the
neutron star kick is not strong enough (as was the case with our
lowered acceleration - MVisc10), the pile-up is insubstantial, and the
kick does not produce an explosion.  In a more borderline case, or if
we follow the collapse further, this slower acceleration may well
drive an explosion.

\section{Implications}

We have presented the results of 6 core-collapse supernova
calculations (a total of 250,000 processor hours) studying the effects
of asymmetric neutrino emission.  If convection occurs without delay
and the explosions are quick, neutrino-driven kicks do not
significantly alter the supernova explosion.  By lowering our
resolution and damping out convection, we are able to produce
``fizzles'' (non-exploding stellar collapse models).  In this
scenario, neutrino-driven kicks are able to seed and drive convection,
ultimately producing an explosion.  The resultant explosions are
asymmetric (Fig. 4).  In the case of our HVisc100 model, the explosion
is more than 6 times more energetic in the direction of the neutron
star's motion.  Note, however, that we only found explosions for very
fast accelerations.  Our lowered acceleration (MVisc10) did not
produce an explosion in the 300\,ms after bounce.  Even so, we believe
a low acceleration version of this neutrino-driven mechanism will
drive an asymmetry for extremely delayed supernova mechanisms, 
e.g. \cite{Bur05}.  But remember that neutrino-driven kicks can 
exist without any outward effect on the supernova explosion if the 
explosion occurs early.

Neutrino-driven kick mechanisms have several distinguishing features
that allow us to differentiate them from ejecta-driven kick
mechanisms.  First, the neutrino driven kicks must occur during the
time of the neutrino emission, that is during the first ten seconds
after the supernova.  As was emphasized by \cite{1998Natur.393..139S},
a kick mechanism of this sort should produce an alignment of the
pulsar velocity with the axis of rotation.  This is because the
components of the kick orthogonal to the axis of rotation average to
nearly zero after many rotations the pulsar makes in the first ten
seconds.  \cite{Joh05} have presented a strong observational evidence of such
an alignment, which appears to lend further support to the neutrino-driven
kick mechanisms.\footnote{We note in passing that no correlation between the
magnitude of the velocity and the strength of the surface magnetic field is
expected from this mechanism \citep{Kus04}.  The $B-v$ correlation is not
expected because the surface magnetic field of a million-year-old pulsar is not
a good representative of the interior magnetic field during the first ten
seconds \citep{dt}.}  \cite{1998Natur.393..139S} have also pointed out that a
kick mechanism of this kind can explain both the proper motions and the rapid
rotations of pulsars.  

It is conceivable, although by no means automatic, that some ejecta-driven
mechanisms could also produce an alignment of the kick with the pulsar's axis
of rotation.  However, by the momentum conservation, the ejecta should recoil
in the direction opposite of the pulsar motion.  If neutrino-driven kicks
help drive the supernova explosion, the explosion ejecta is strongest in the
same direction of the motion of the neutron star.  This means that there
should be more mixing in the direction of the neutron star's motion for these
neutrino-driven kicks, and elements like nickel will mix further out in
this direction \citep{Hun05}.  If one believes the compact object
identification of Sgr A East \citep{Par05}, this extended mixing in
the direction of the neutron star kick has already been observed.
This observation is circumstantial, but it does show the possibility
of distinguishing between these two kick mechanisms.  We stress that 
this is only valid if the neutrino-driven mechanism is responsible for 
the supernova explosion (it is not valid if the explosion is quick).

An additional independent confirmation of the neutrino-driven kicks can come
from observations of the gravity waves.  It is well-known
that a departure from spherical symmetry is necessary for generating
gravitational waves.  A neutron star emitting neutrinos anisotropically, while
rotating around some axis that is not aligned with the direction of the 
anisotropy, creates a source of gravity waves observable by Advanced LIGO and
LISA in the event of a nearby supernova \citep{2004PhRvD..69b4008L}.  The
signal discussed by \cite{2004PhRvD..69b4008L} is caused by the rotating ray of
overdensity in the neutrino distribution.  Gravity waves may also be
sourced by anisotropic distribution of oscillating neutrinos deep inside the
neutron star \citep{2002PhRvD..65f1503C}. 

\section{Conclusions}

To summarize, we have shown that pulsar kick mechanisms based on anisotropic
emission of neutrinos or other weakly interacting particles from the cooling
neutron star can increase the energy gained by the shock, hence improving the
prospects for a successful explosion.  A distinguishing feature of this class
of mechanisms is asymmetric explosion enhanced in the direction of the motion
of the neutron star.   

\acknowledgements{The work of C.L.F. was funded in part under the
auspices of the U.S.\ Dept.\ of Energy, and supported by its contract
W-7405-ENG-36 to Los Alamos National Laboratory, and by a DOE SciDAC
grant DE-FC02-01ER41176.  The work of A.K. was supported in part by by
the US Department of Energy grant DE-FG03-91ER40662 and by NASA grants
ATP02-0000-0151 and ATP03-0000-0057.}

{}

\begin{deluxetable}{lcccc}
\tablewidth{0pt}
\tablecaption{Simulation Parameters}
\tablehead{
  \colhead{Model}
& \colhead{Particle}
& \colhead{EOS\tablenotemark{a}}
& \colhead{$\alpha,\beta$\tablenotemark{b}}
& \colhead{$a_{\rm Kick}$} \\

\colhead{}
& \colhead{Number}
& \colhead{}
& \colhead{}
& \colhead{($10^7 {\rm cm \, s^{-2}}$)}
}
\startdata
StanHR0 & 399150 & Her & 1.5,3.0 & 0 \\
StanHR100 & 399150 & Her & 1.5,3.0 & 100 \\
Stan0 & 101685 & Her & 1.5,3.0 & 0 \\
MVisc10 & 101685 & LS & 2.0,4.0 & 10 \\ 
HVisc0 & 101685 & LS & 3.0,6.0 & 0 \\
HVisc100 & 101685 & LS & 3.0,6.0 & 100 \\

\enddata
\tablenotetext{a}{Her:  The couple equation of state described 
in \cite{Her94}.  LS:  The equation of state using the \cite{Lat91} 
equation of state down to densities of $10^9 {\rm g \, cm^{-3}}$.}
\tablenotetext{b}{$\alpha$ and $\beta$ correspond to the standard 
SPH representation of the bulk and von-Neuman-Richtmyer viscosities 
respectively \citep{FRW05}.}

\end{deluxetable}
\clearpage

\begin{deluxetable}{lccccc}
\tablewidth{0pt}
\tablecaption{Simulations}
\tablehead{
  \colhead{Model}
& \colhead{$t_{\rm Bounce}$\tablenotemark{a}}
& \colhead{$\rho_{\rm Bounce}$\tablenotemark{b}}
& \colhead{$t_{\rm exp}$\tablenotemark{c}}
& \colhead{$E_{\rm exp,x>0}/E_{\rm exp,x<0}$\tablenotemark{d}}
& \colhead{$p_x,p_y,p_z$\tablenotemark{e}} \\

\colhead{}
& \colhead{(ms)}
& \colhead{($10^{14}{\rm g cm^{-3}}$)}
& \colhead{(ms)}
& \colhead{}
& \colhead{($10^{40} {\rm g cm s^{-1}}$)}
}
\startdata

StanHR0 & 181 & 3.56 & 40 & 1.02 & -0.04,-0.09,-0.36 \\ 
StanHR100 & 178 & 3.42 & 40 & 1.00 & 0.08,0.12,0.36 \\ 
Stan0 & 248 & 3.87 & N/A & N/A & N/A \\ 
MVisc10 & 298 & 4.04 & $>300$ & N/A & N/A \\ 
HVisc0 & 208 & 3.66 & N/A & N/A & N/A \\
HVisc100 & 208 & 3.86 & 90 & 6.54 & 5.4,0.96,0.11 \\

\enddata
\tablenotetext{a}{Time between the onset of collapse of the 
S157B2 progenitor to bounce.}
\tablenotetext{b}{Core density at bounce.}
\tablenotetext{c}{Time after bounce when the explosion shock has 
reached 500\,km.  N/A refers to models that do not explode.}
\tablenotetext{d}{Ratio of explosion energies for all particles with 
$x>0$ divided by those particles with $x<0$.  The explosion energy 
is defined as the kinetic energy of those particles with radial 
velocities greater than 0.}
\tablenotetext{e}{Total momentum of the particles with density 
less than $10^{12} {\rm g cm^{-3}}$.}
\end{deluxetable}
\clearpage

\newpage
\begin{figure}
\plottwo{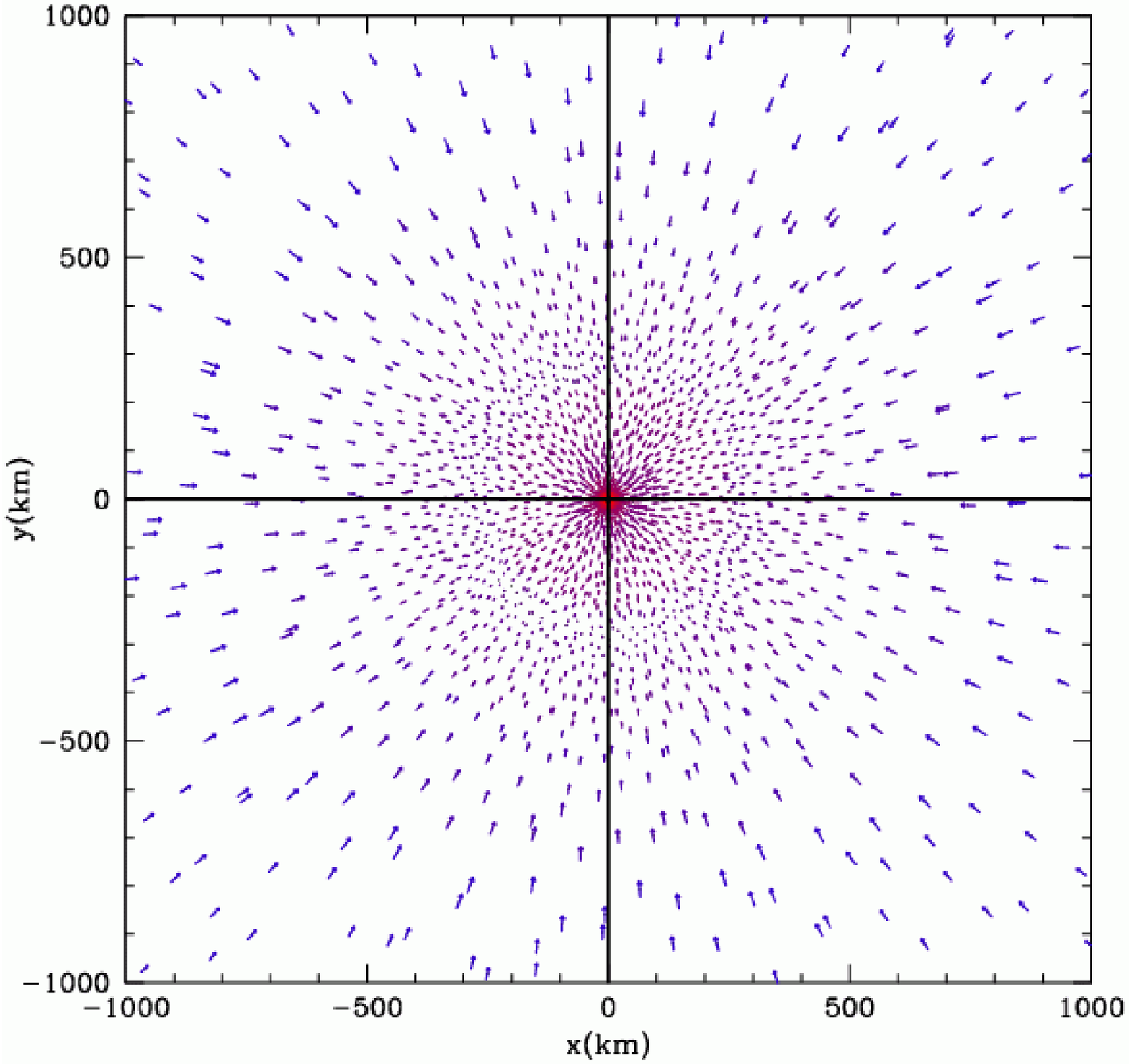}{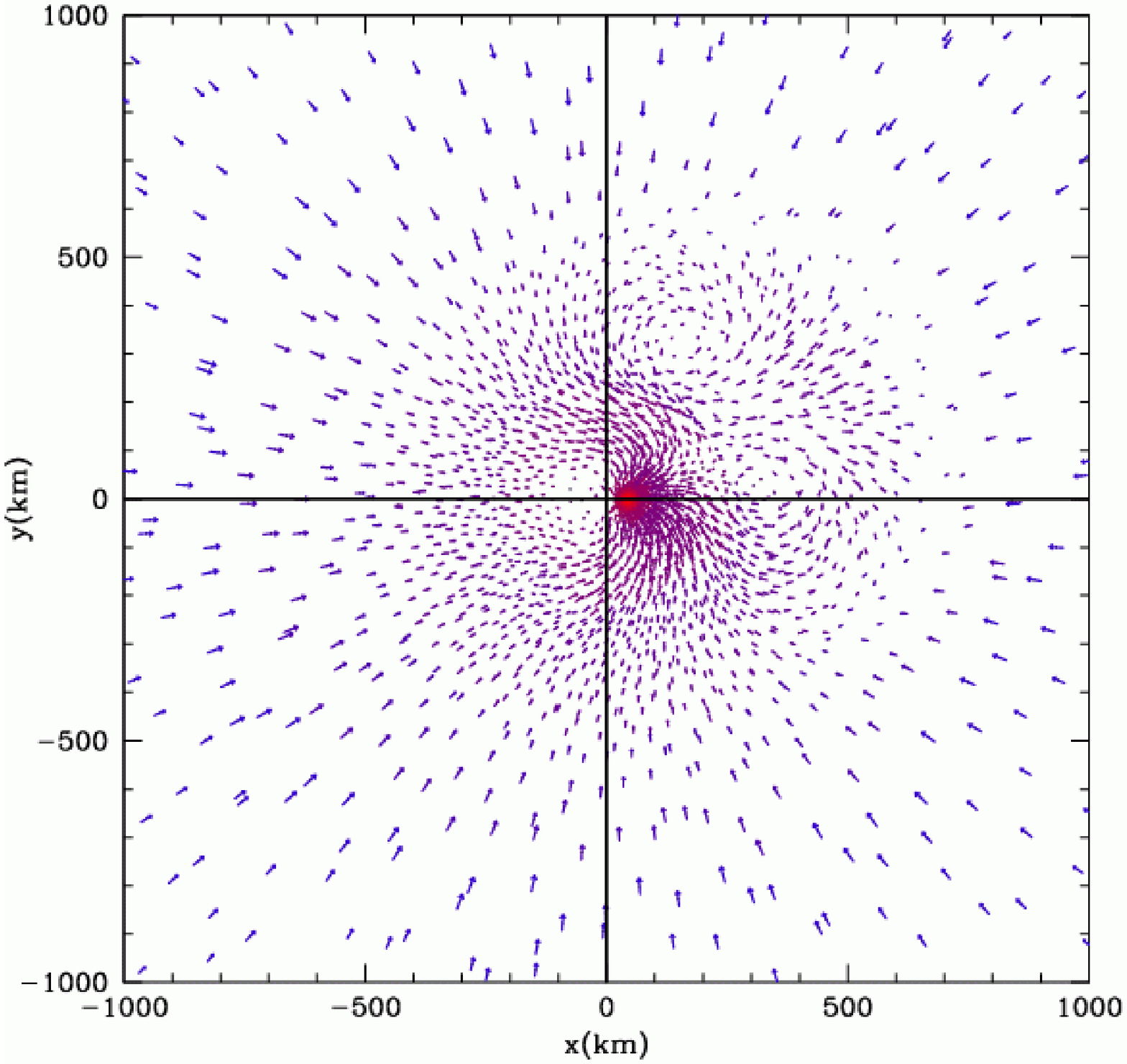}
\caption{Comparison of the symmetric ``HVisc0'' (left) and kick
``HVisc100'' (right) models 90\,ms after bounce.  These plots show a
slice of data centered on the z=0 plane with the kick in the positive
x direction.  Shading denotes entropy (dark is low, light is high) and
the direction and length of arrows denote the direction and magnitude
of the velocity.  Note that the kicked model has developed some strong
convection which is pushing out the accretion shock.  It ultimately
developes into a strong explosion (Fig. 3).  Primarily because of the 
low resolution, convection does not develop in the symmetric model and 
this model does not explode.} 
\label{fig:xycomp}
\end{figure}
\clearpage

\newpage
\begin{figure}
\plotone{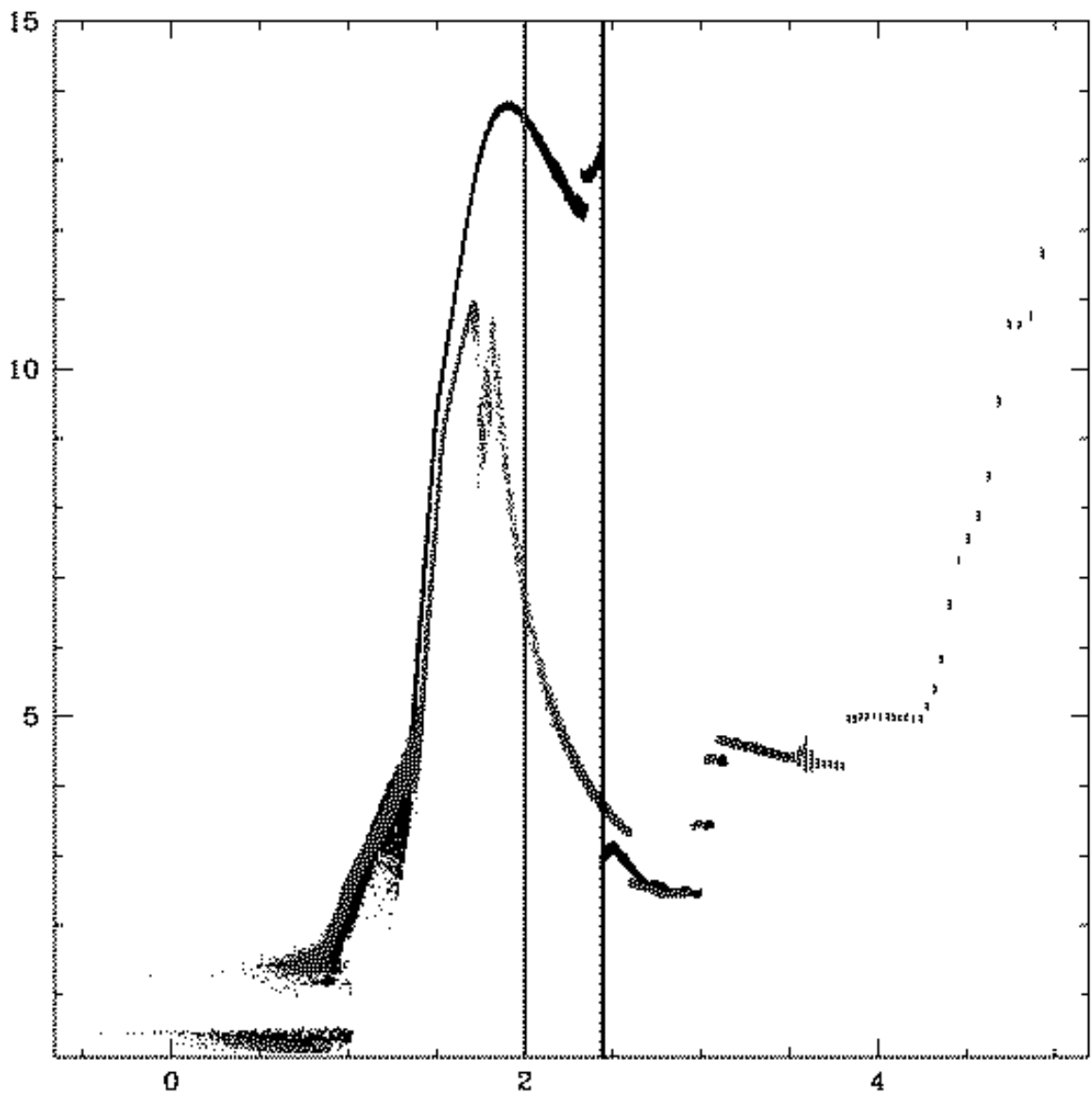}
\caption{Entropy versus radius 40\,ms after bounce for models using
the coupled equation of state from \cite{Her94} (light particles)
compared with those using the Lattimer-Swesty \citep{Lat91} equation
of state down to low densities (dark particles).  The vertical lines
correspond to the positions of the accretion shock for these two
models.  Note that although the entropy from the coupled equation of
state is lower than that using Lattimer-Swesty down to low densities,
the entropy gradient out to the shock is much higher.  This is more
conducive to convection.  The entropy gradient across the shock is
also much more gradual in the case of the coupled equation of state.  
This is because nuclear dissociation and burning is playing a strong 
role in determining the entropy.}
\label{fig:svsr}
\end{figure}
\clearpage

\newpage
\begin{figure}
\plottwo{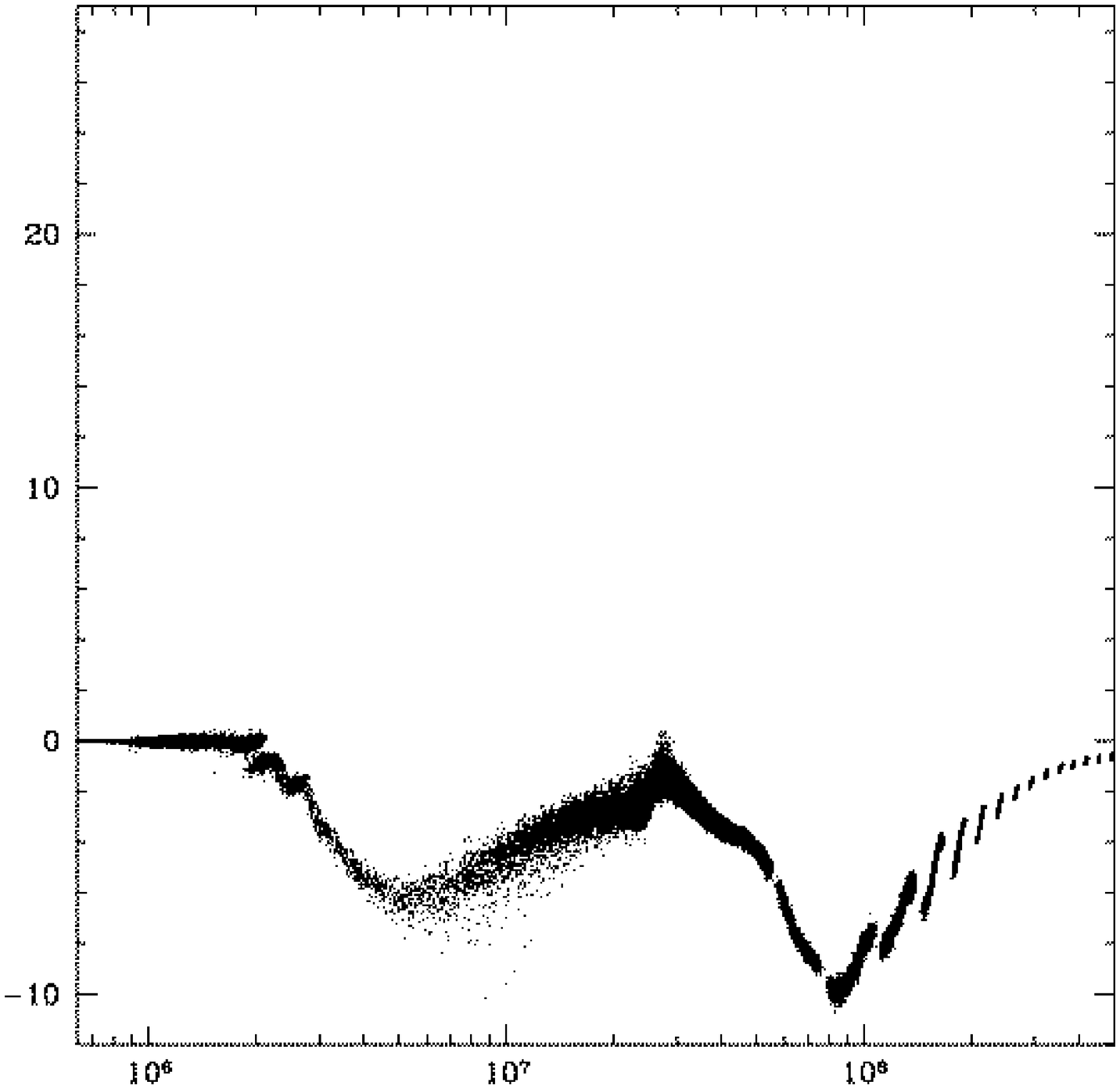}{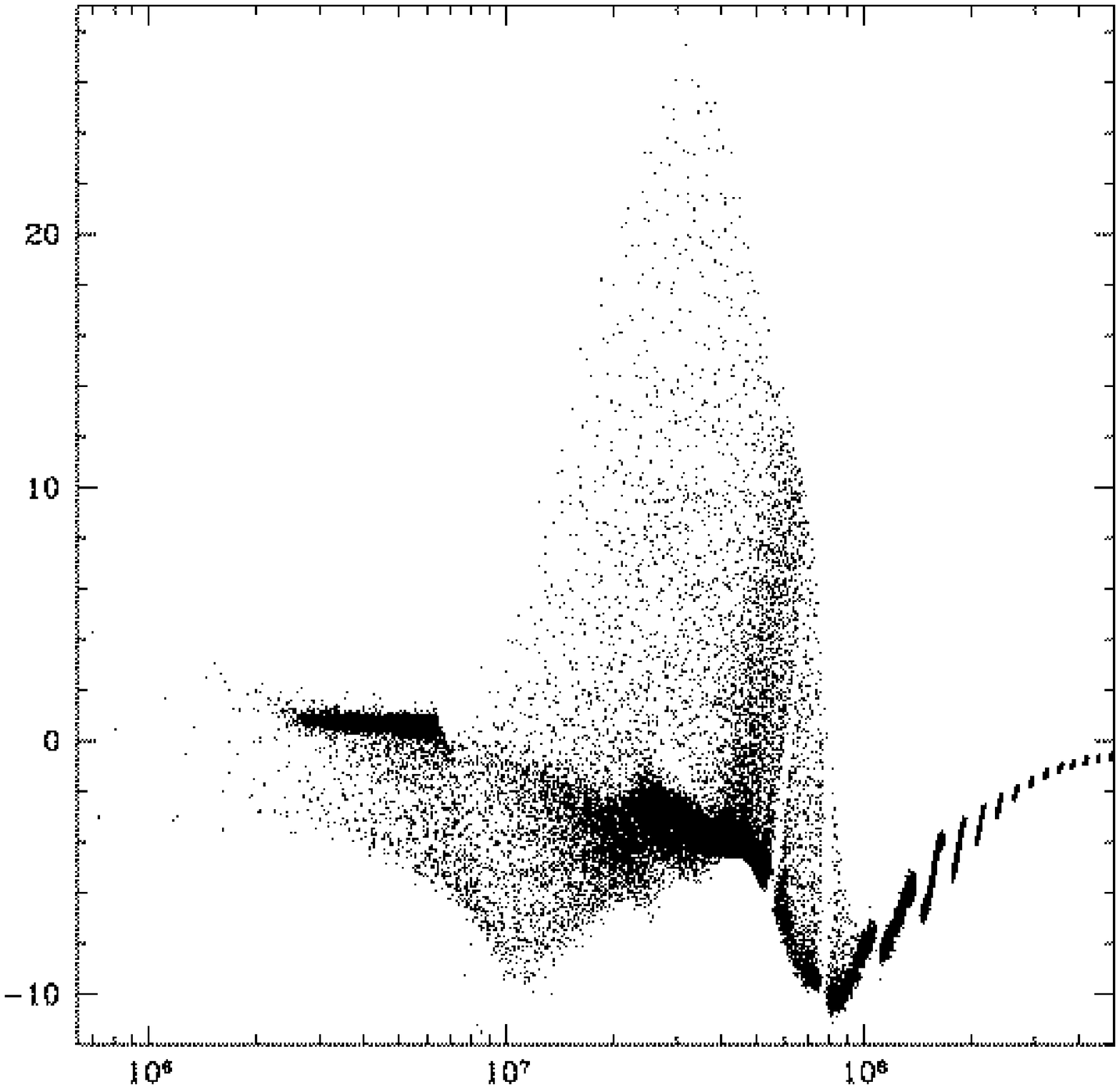}
\caption{Radial velocity versus radius for our symmetric (HVisc0) and 
kicked (HVisc100) models, showing clearly the lack of convection in 
the symmetric simulation in stark contrast to the convection that has 
developed in the kicked model.  For all practical purposes, the symmetric 
model is behaving as one might expect in a 1-dimensional simulation.}
\label{fig:vrvr}
\end{figure}
\clearpage

\newpage
\begin{figure}
\plotone{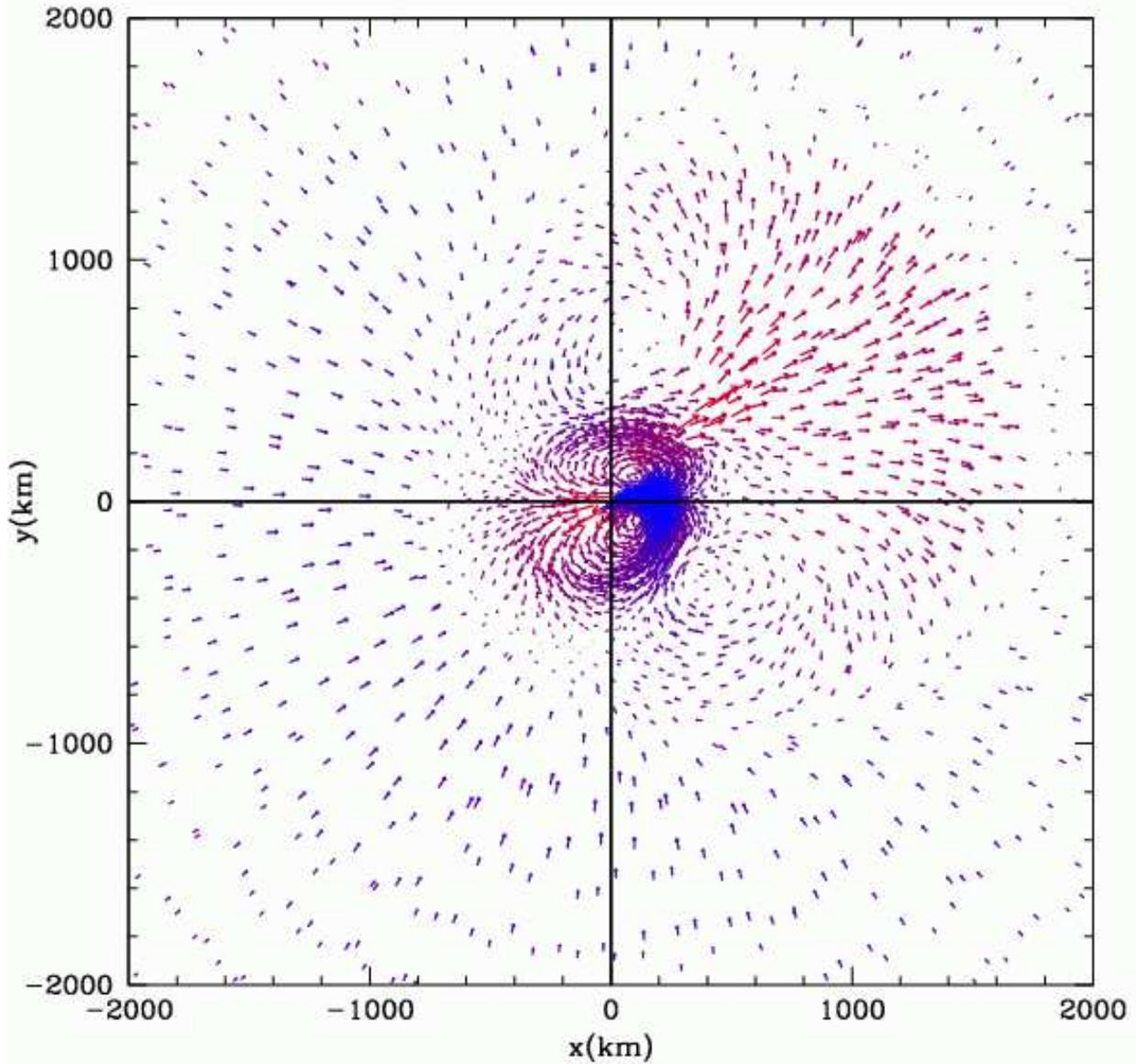}
\caption{Slice of the kicked simulation ``HVisc100'' 160\,ms after
bounce.  An explosion has been launched and, where it is strongest,
has now nearly reached 2000\,km.  The explosion ejecta is strongest in
roughly the {\it same} direction as the neutron star.  Because the
explosion ejecta is driven by convection, and the seeds for this
convection are building on small asymmetries in the collapsing core,
the fastest moving ejecta is not exactly aligned with the motion of
the kick.  Ejecta driven kicks predict the exact opposite - the ejecta
moves in the opposite direction of the neutron star.}
\label{fig:xyfinal}
\end{figure}
\clearpage

\newpage
\begin{figure}
\plotone{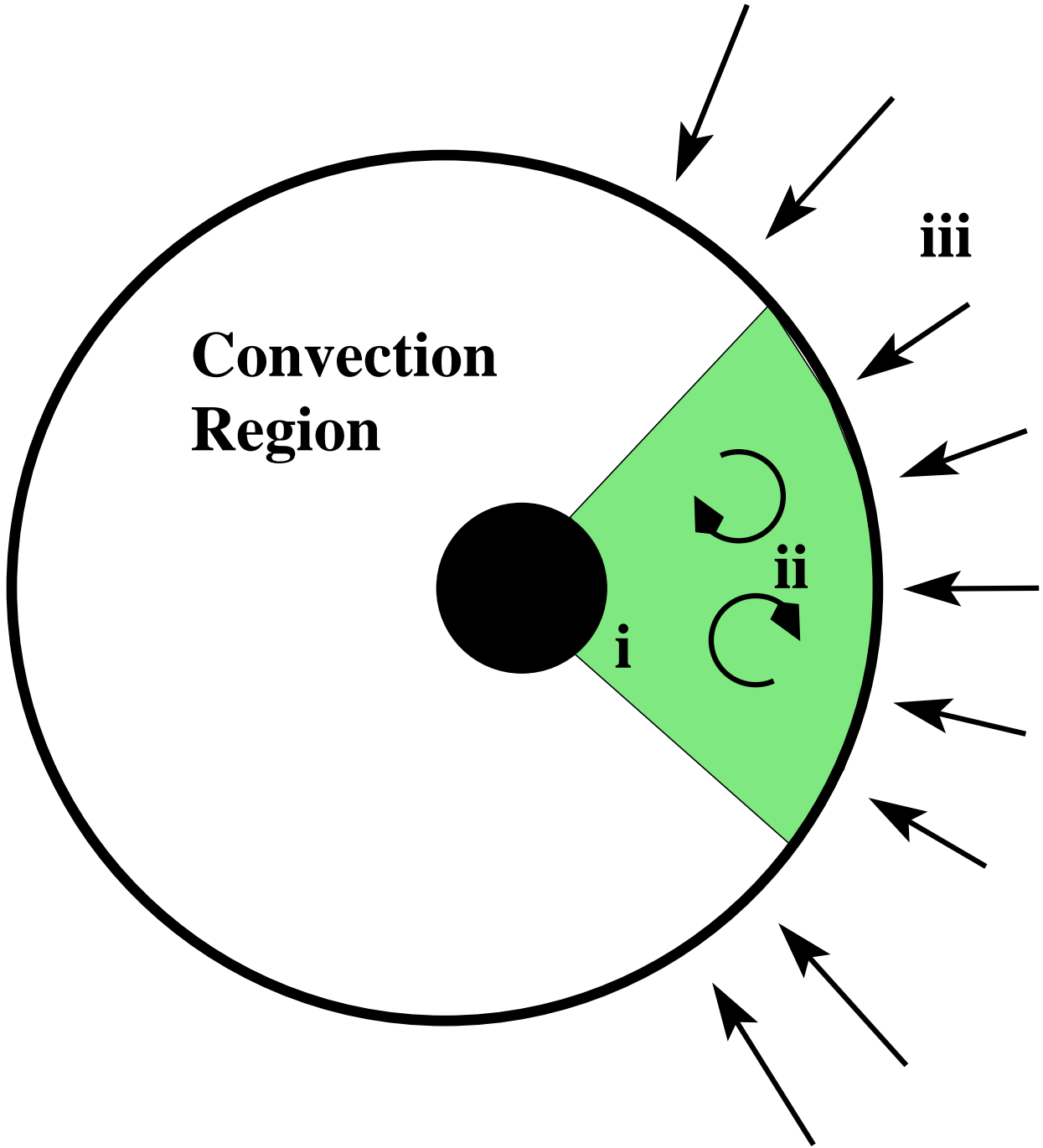}
\caption{Possible causes for the ejecta asymmetry: i) asymmetric
neutrino heating or ram-heating caused by the motion of the
proto-neutron star, ii) weakened pressure at the accretion shock, or
iii) convection seeded by the motion of the proto-neutron star
(possibly caused by minor effects of the previous two effects).  Our
analysis suggests that this latter cause is indeed the cause of the
ejecta asymmetry in our exploding models.}
\label{fig:diag}
\end{figure}
\clearpage

\newpage
\begin{figure}
\plottwo{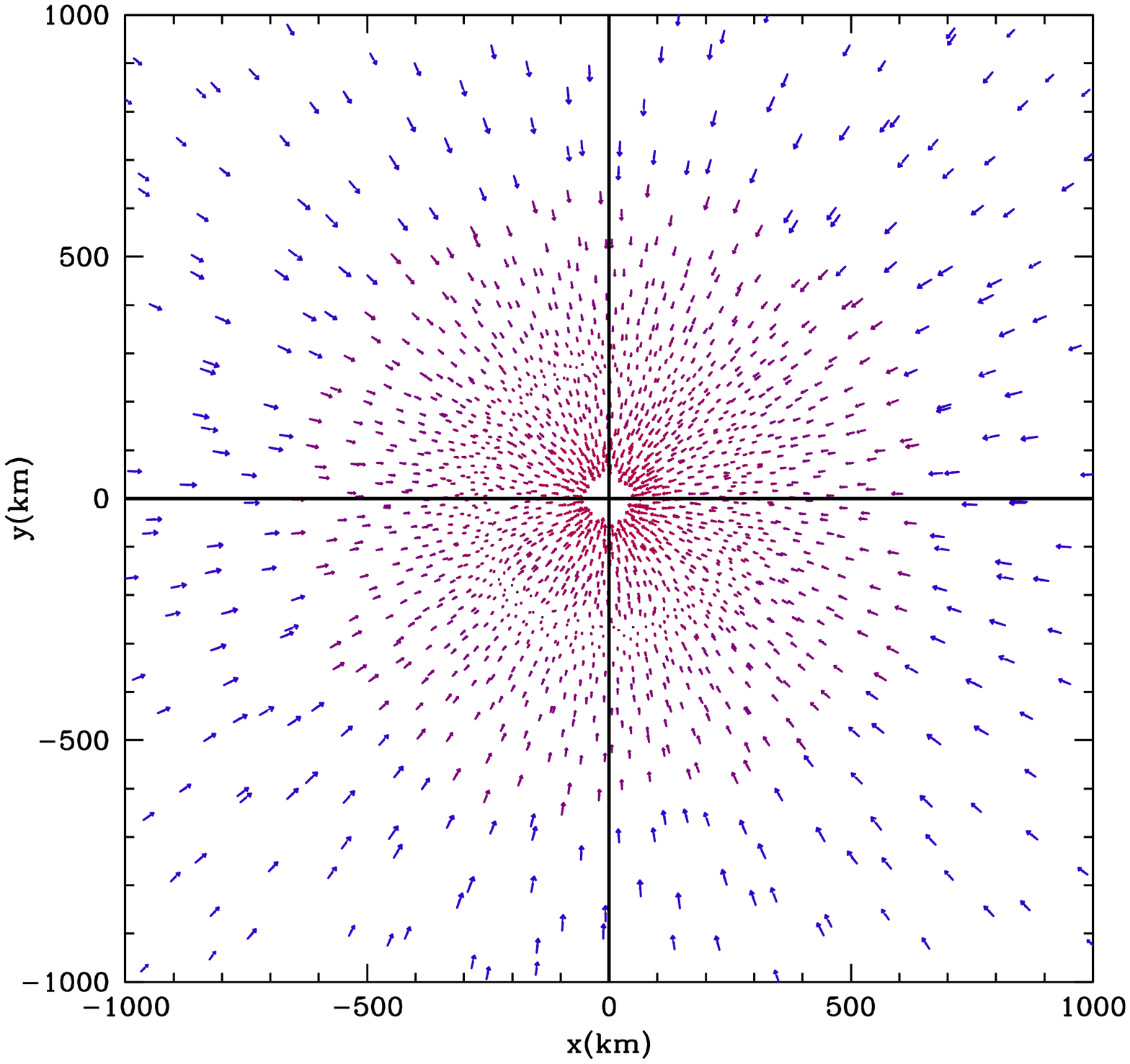}{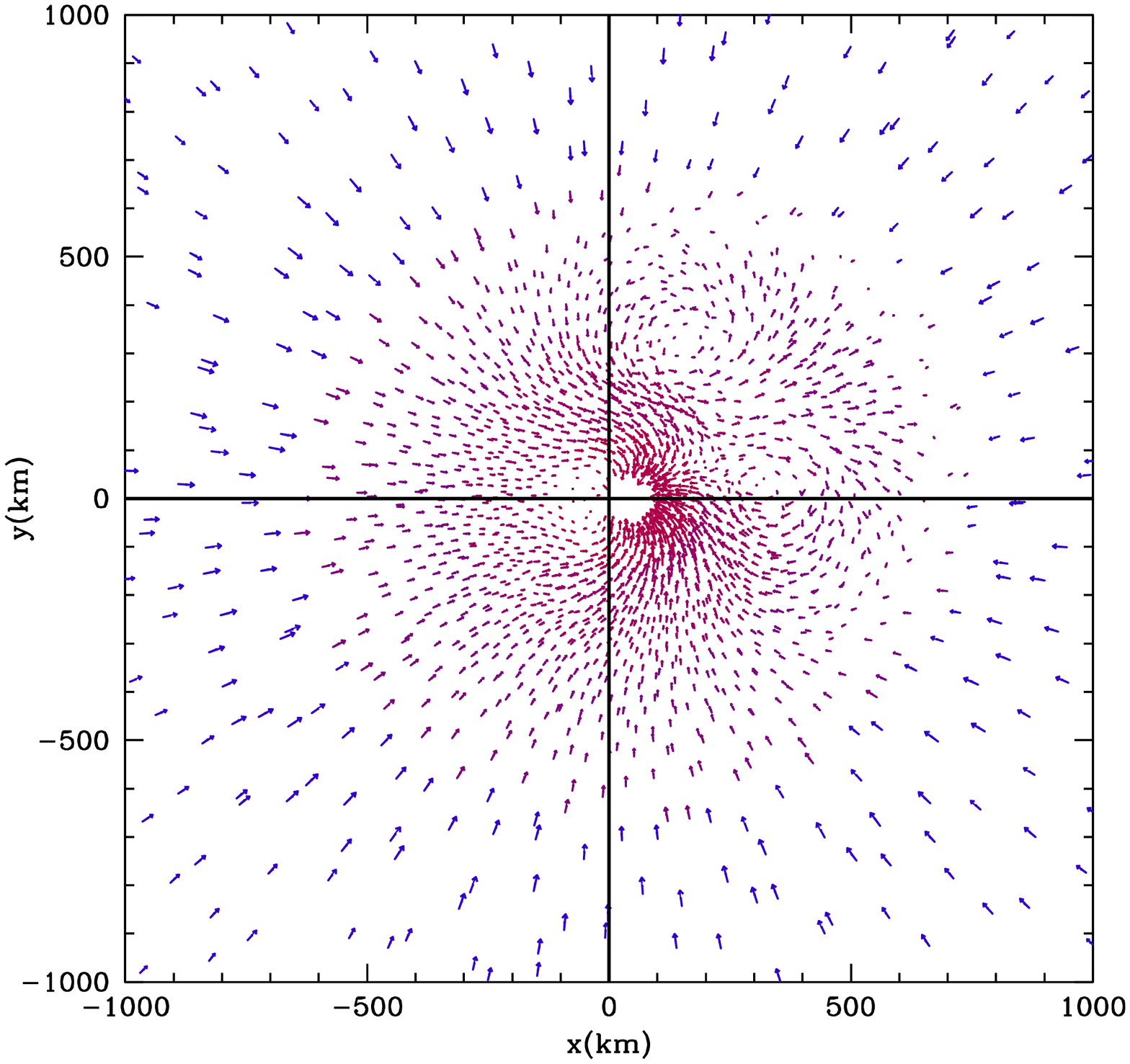}
\caption{Slices of our symmetric (HVisc0) and kicked (HVisc100) models 
colored by the heating from neutrinos.  The symmetric model exhibits no 
asymmetry in its net heating.  Although there is more heating 
in the positive x-direction for the asymmetric model, if we use the 
center of our neutron star as the zero point, the neutrino heating 
for particles with $x-x_{\rm NS}>0$ equals that of the particles with 
$x-x_{\rm NS}<0$.}
\label{fig:neut}
\end{figure}
\clearpage

\newpage
\begin{figure}
\plotone{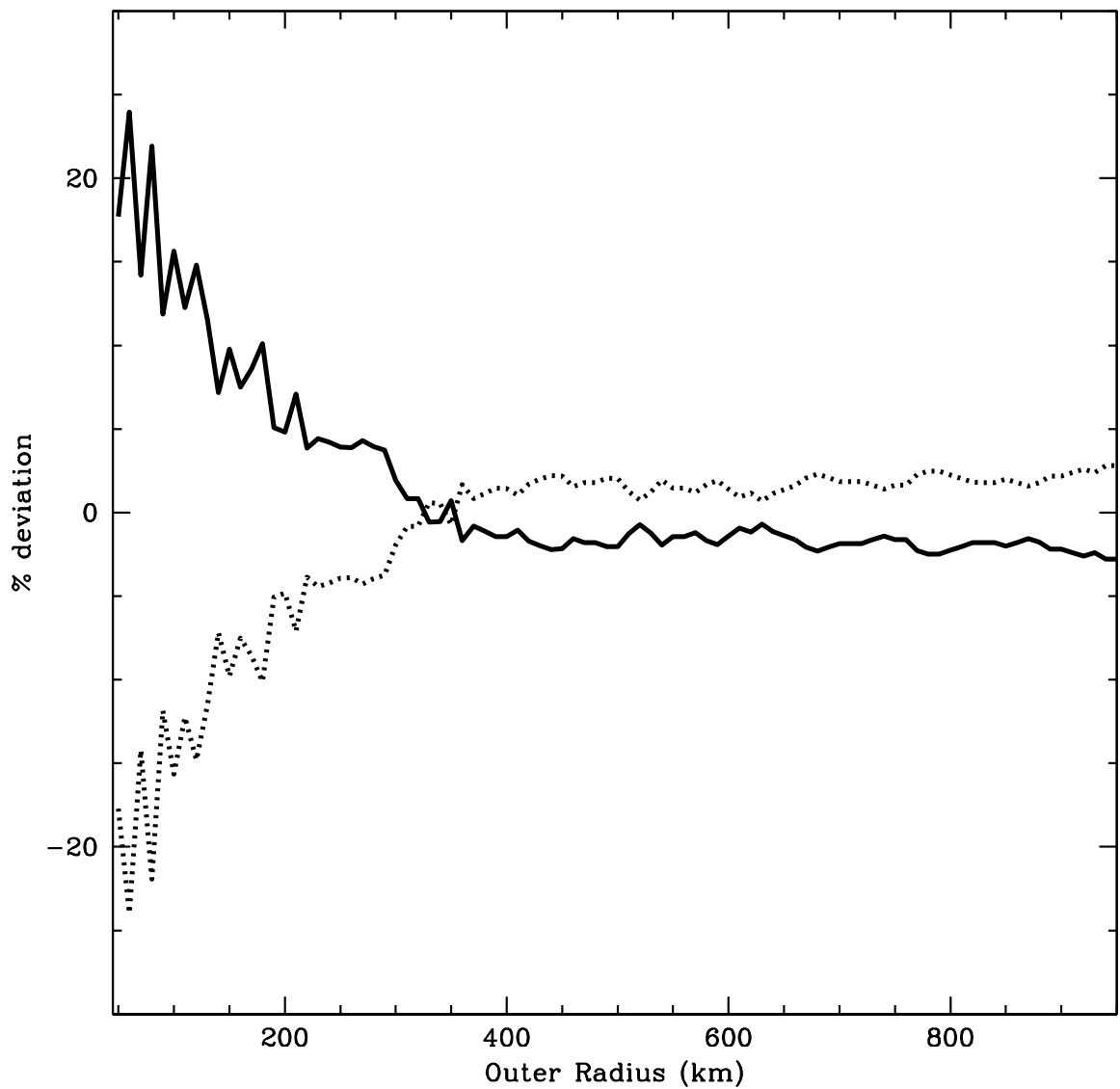}
\caption{Percentage deviation from the mean of the mass as a function
of radial bins (where the zero point in the radius is determined by
the center of the neutron star) 75\,ms after bounce.  The solid line
denotes the mass in a 15$^{\circ}$ cone leading the neutron star, the
dotted line denotes the mass of a similar cone trailing the neutron
star.  Out to roughly 300\,km, there is more mass ahead of the neutron
star than the mean as it piles up against the neutron star.  In the
wake, the mass is below the mean.  This mass increase is nearly 20\%
near the neutron star's surface.  Beyond 300\,km, the fact that the
center of the neutron star is off of the center of the collapsing star
means that the density ahead of the neutron star is lower, and the
total mass is lower.}
\label{fig:mass}
\end{figure}
\clearpage

\newpage
\begin{figure}
\plotone{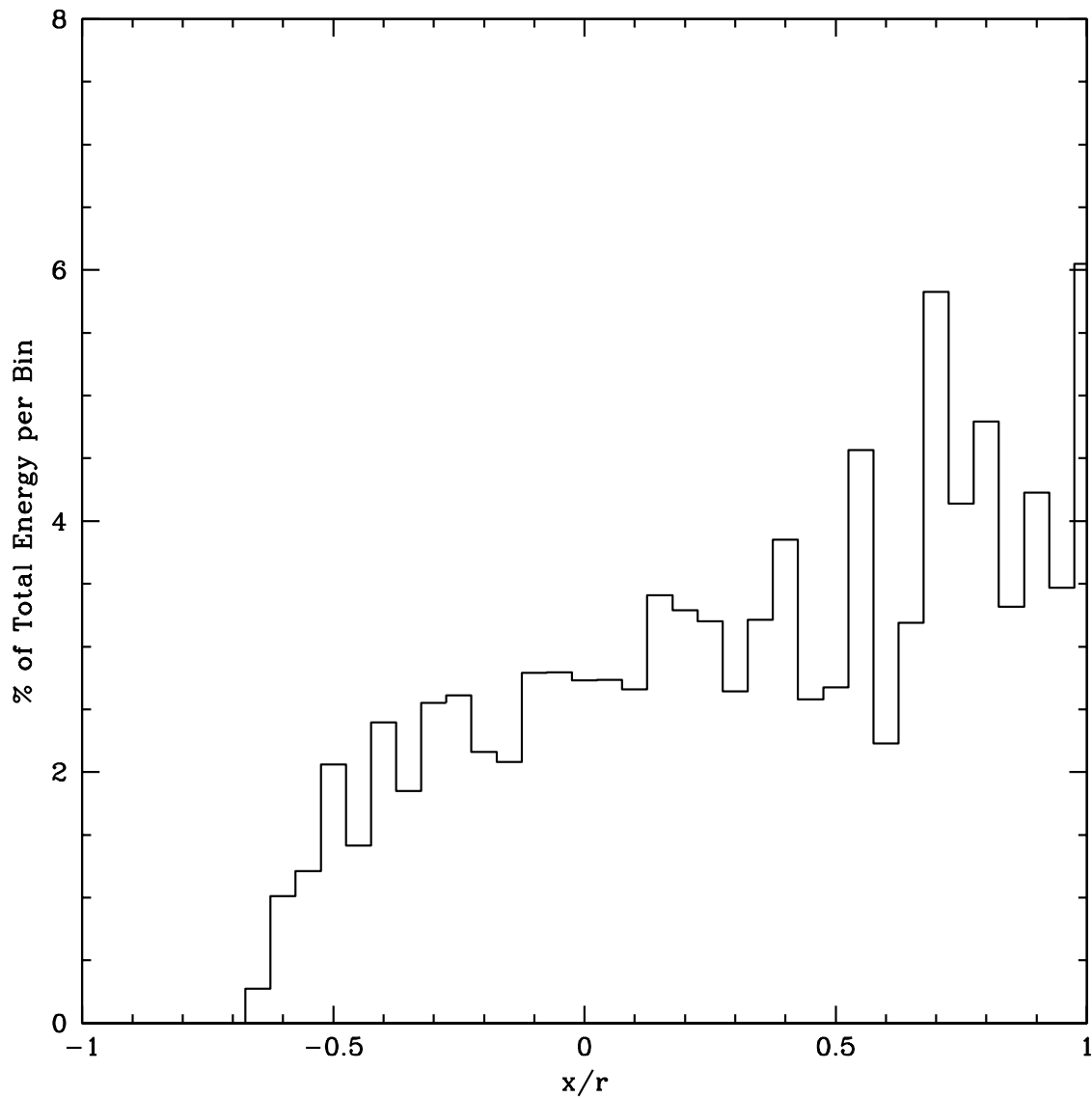}
\caption{Percent of total energy per angular bin as a function of
angle (x position over radius) for mass 70-110\,km from the center of
the proto-neutron star 75\,ms after bounce.  In this plot we see just
how important the mass pile-up shown in Figure 7 is for the energy in
this region.  70\% of the total energy is ahead of the moving neutron
star (the neutron star is moving in the positive x-direction), peaking
directly ahead of the neutron star's motion.}
\label{fig:energy}
\end{figure}
\clearpage

\end{document}